\documentclass{article}
\usepackage{pack}
\usepackage{amsfonts,amsmath,amssymb,graphicx}
\begin{document}
% Numbering equations by section
\def\theequation{\thesection.\arabic{equation}}
\makeatletter
\@addtoreset{equation}{section}
\makeatother
% Numbering figures by section
\def\thefigure{\thesection.\arabic{figure}}

%Proof environment
\newenvironment{proof}[1][Proof]{\textbf{#1.} }{\hfill \rule{0.5em}{0.5em}}
% Miscellaneous
\newcommand\egaldef{\stackrel{\mbox{\upshape\tiny def}}{=}}

% Next two lines for the indicator function
\newcommand\1{\leavevmode\hbox{\rm \small1\kern-0.35em\normalsize1}}
\newcommand\ind[1]{\1_{\{#1\}}}        

\newcommand{\lra}{\mathop{\leftrightarrows}}
\newcommand{\rla}{\mathop{\rightleftarrows}}
\newcommand{\ra}{\mathop{\rightarrow}}

\newcommand\EE{\mathsf{E}}
\newcommand{\be}{\begin{equation}}
\newcommand{\ee}{\end{equation}}
\newcommand{\bea}{\begin{eqnarray}}
\newcommand{\eea}{\end{eqnarray}}
\def\DD{\displaystyle}

%%%% Please replace the question marks by title, short title (for running page
%%%% headers), and authors' names, respectively.

%%%%
\renewcommand{\titre}{Stochastic Deformations of Sample Paths of
Random Walks and Exclusion Models}
\renewcommand{\smalltitre}{Stochastic deformations of sample paths}
\renewcommand{\auts}{Guy Fayolle, Cyril Furtlehner} 
\tit

\begin{abstract}
This study in centered on models accounting for stochastic
deformations of sample paths of random walks, embedded either in
$\mathbb{Z}^2$ or in $\mathbb{Z}^3$. These models are immersed in
multi-type particle systems with exclusion. Starting from examples, we
give necessary and sufficient conditions for the underlying Markov
processes to be reversible, in which case their invariant measure has
a Gibbs form. Letting the size of the sample path increase, we find
the convenient scalings bringing to light phase transition phenomena.
Stable and metastable configurations are bound to time-periods of
limiting deterministic trajectories which are solution of nonlinear
differential systems: in the example of the ABC model, a system of
Lotka-Volterra class is obtained, and the periods involve elliptic,
hyper-elliptic or more general functions. Lastly, we discuss briefly the
contour of a general approach allowing to tackle the transient regime via
differential equations of Burgers' type.
\end{abstract}

%%%% Please provide some keywords
\index{random walk}
\index{exclusion}
\index{thermodynamic limit}
\index{Burger's equation}
\index{phase transition}
%%%%%%%%%%%%%%%%%%%%%%%%%%%%%%%%%%%%%
%% Please put here the text of your paper %%
%%%%%%%%%%%%%%%%%%%%%%%%%%%%%%%%%%%%%

\section{Introduction}\label{INTRO}
We are interested in models describing evolution of sample paths of
random walks, when they are submitted to random local deformations
involving possibly several links. Roughly speaking, given a finite
sample path, say of size $N$, forming a not necessarily closed curve,
the problem will be to characterize the evolution of an associated
family $\{\mathbb{Y}_i,i=1,\ldots,N\}$ of Markov processes in the
thermodynamic limit as $N\to\infty$. This requires to guess and to
find  the interesting scalings.

\medskip\noindent
In a previous study \cite{FaFu}, we considered random walks on a
square lattice, deformations involved pairs of links and occurred at
the epoches of Poisson jump processes in continuous time (see section
\ref{SEC1} for a more exact definition). The analysis was carried out
by means of an explicit mapping, which led to view the system as a
coupling of two exclusion processes. 

Starting from a number of observations, we intend to hint in this
paper that the model in \cite{FaFu} can indeed be cast into a broader
class, the ultimate goal being to propose methods of wide
applicability concerning the following questions:
\begin{itemize}
\item conditions ensuring Gibbs states and explicit forms of the
corresponding invariant measures;
\item steady-state equations in the thermodynamic limit as
$N\to\infty$, and their solutions in the case of Gibbs states, but
also in situations involving permanent currents;
\item hydrodynamic and transient equations, when $N$ is sufficiently
large, yielding thus a complete picture of the evolution.
\end{itemize}

Generalizations of the model in $\mathbb{Z}^2$ can follow two natural
trends. First, in modifying the construction of the random walk.
Indeed, in the square, lattice we dealt with a $4$-letter alphabet.
Considering instead a finite alphabet of $l$ letters is then
tantamount to constructing random walks with oriented links, whose angular
affixes are multiples of $\frac{2k\pi}{l},k=0,\ldots,l-1$. The case
$l=2$ corresponds to the simple exclusion process in $\mathbb{Z}$, and
$l=3$ yields the so-called \emph{ABC model}. 

\medskip Another possible extension is to relax the constraint that the walk
lives in $\mathbb{Z}^2$ and to define a stochastic deformation process
in higher dimension. In the sequel, we shall restrict ourselves to some paradigms in
$\mathbb{Z}^2$ and $\mathbb{Z}^3$. 

\medskip In section \ref{SEC1}, we define a
class of two-dimensional models, together with related patterns in
$\mathbb{Z}^3$, in terms of exclusion particle systems. Section
\ref{SEC2} is devoted to stochastic reversibility of the Markov
processes of interest and to the Gibbs form of their invariant
measure. In section \ref{SEC3}, the non-symmetric classical ABC model
is solved (fundamental scaling, phase transitions, classification of
stable configurations) through the analysis of a Lotka-Volterra
differential system. The concluding section \ref{CONC} gives a brief
overview of ongoing research about large scale dynamics,
nonequilibrium and transient regimes.

%%%%%%%%%%%%%%%%%%%%%%%%%%%%%%%%%%%%
\section{Model descriptions via exclusion particle systems}\label{SEC1}
Our main objective in this section is to show how the evolution of the
sample paths of random walks can be fruitfully described by means of
particle exclusion processes.

Beforehand, to avoid repetition and clumsy notation, let us emphasize
that we shall only deal with jump Markov processes in continuous time.
So implicitly the word \emph{transition rate} will always refer to
some underlying generator. Also, $N$ will always stand for the size of
the sample path, or equivalently the number of its links.
 
 \subsection{Preliminaries}
In $\mathbb{Z}$, the simple exclusion model coincides with the well
known KPZ system (see e.g. \cite{KPZ}), which represents a fluctuating
and eventually growing interface. This system is coded by a sequence
of binary variables $\{\tau_j\}$, $j=1,\ldots, N$, depending on
whether a particle is present or not, with asymmetric jump rates. This
system has been extensively studied. In particular, the invariant
measure has been obtained in a closed matrix form solution, for fairly
arbitrary parameters and boundary conditions \cite{DeEvHaPa}. Large
scale dynamics has also been analyzed \cite{Sp}, showing Burgers'
equations \cite{Bu}.

%%%%%%%%%%%%%%%
\subsection{$2$-dimensional models}\label{MODELS}
\emph{1) The triangular lattice and the ABC model.} \   Here the evolution of the random walk is restricted to the triangular lattice. Each link (or step) of the walk is either $1$, $e^{2i\pi/3}$  or $e^{4i\pi/3}$, and quite naturally will be said to be of type A, B
 and C, respectively. This corresponds to the so-called \emph{ABC
 model}, since there is a coding by a $3$-letter alphabet. The set of
 \emph{transitions} (or reactions) is given by
\begin{eqnarray}
AB\ \lra_{p^+}^{p^-}\ BA, \qquad 
BC\ \lra_{q^+}^{q^-}\ CB, \qquad 
CA\ \lra_{r^+}^{r^-}\ AC, \qquad 
\end{eqnarray}
where the introduced rates are fixed, but not necessarily equal. Also we impose \emph{periodic
boundary conditions} on the sample paths. This model was first
introduced in \cite{EvFoGoMu} in the context of particles with
exclusion, and a Gibbs form corresponding to reversibility 
has been found in \cite{EvKaKoMu} in some cases.

\medskip
\emph {2)  The square lattice and coupled exclusion model.} \  This model was introduced in \cite{FaFu} to analyze stochastic
deformations of a walk in the square lattice, and it will be referred to
from now on as the $\{\tau_a\tau_b\}$ model. Assuming links
counterclockwise oriented, the following  transitions can take place.
\begin{eqnarray}
AB\ \rla^{\lambda^{ab}}_{\lambda^{ba}}\ BA, \qquad 
BC\ \rla^{\lambda^{bc}}_{\lambda^{cb}}\ CB, &\qquad& 
CD\ \rla^{\lambda^{cd}}_{\lambda^{dc}}\ DC, \qquad 
DA\ \rla^{\lambda^{da}}_{\lambda^{ad}}\ AD,\nonumber \\[0.2cm] 
AC\ \rla^{\gamma^{ac}}_{\delta^{bd}}\ BD, \qquad 
BD\ \rla^{\gamma^{bd}}_{\delta^{ca}}\ CA, &\qquad& 
CA\ \rla^{\gamma^{ca}}_{\delta^{db}}\ DB, \qquad 
DB\ \rla^{\gamma^{db}}_{\delta^{ac}}\ AC.\nonumber
\end{eqnarray}
We studied a rotation invariant version of this model when
\[
\begin{cases}
\lambda^+ \egaldef \lambda^{ab} = \lambda^{bc} = \lambda^{cd} =
 \lambda^{da}, \\ 
\lambda^- \egaldef \lambda^{ba} = \lambda^{cb} =
 \lambda^{dc} = \lambda^{ad}, \\ 
\gamma^+ \egaldef \gamma^{ac} =
 \gamma^{bd} = \gamma^{ca} = \gamma^{db}. \\ 
\gamma^- \egaldef \delta^{ac} = \delta^{bd} = \delta^{ca} = \delta^{db}.
\end{cases}
\]
Define the  mapping 
$(A,B,C,D)\to(\tau^a,\tau^b)\in\{0,1\}^2$, 
such that
\[
\begin{cases}
A \to (0,0),\\
B \to (1,0),\\
C \to (1,1),\\
D \to (0,1).\\
\end{cases}
\]
Then the dynamics can be formulated in terms of coupled exclusion
 processes. The evolution of the sample path is represented by a
 Markov process with state space the $2N$ binary random variables
 $\{\tau_j^a\}$ and $\{\tau_j^b\}$, $j=1,\ldots, N$, taking value $1$
 if a particle is present and $0$ otherwise. The jump rates to the
 right ($+$) or to the left ($-$) are given by
\begin{equation}\label{taux}
\begin{cases}
\lambda_a^{\pm}(i)={\bar \tau_i^b}{\bar
\tau_{i+1}^b}\lambda^{\mp}+\tau_i^b \tau_{i+1}^b\lambda^{\pm} + {\bar
\tau_i^b}\tau_{i+1}^b\gamma^{\mp} + \tau_i^b{\bar
\tau_{i+1}^b}\gamma^{\pm},\\[0.2cm] \lambda_b^{\pm}(i)={\bar
\tau_i^a}{\bar \tau_{i+1}^a}\lambda^{\pm}+\tau_i^a
\tau_{i+1}^a\lambda^{\mp} + {\bar \tau_i^a}\tau_{i+1}^a\gamma^{\pm} +
\tau_i^a{\bar \tau_{i+1}^a}\gamma^{\mp}.
\end{cases}
\end{equation}
Notably, one sees the jump rates of a given sequence are
locally conditionally defined by the complementary sequence.

\medskip
\emph{3) An extended stochastic clock model.}  \  We propose an extension of the preceding model for an arbitrary
two-dimensional regular graph. To this end, consider a random walk
composed of oriented links, the affixes of which take values
$\omega_k=\exp(2ik\pi/n),k=1,\ldots,n$, the $n$-th roots of unity.
There are two different situations.

\medskip
\  \emph{(a)} $n=2p+1$ is odd. Then the walk cannot have a \emph{fold} of
two successive links, so that local displacements of edges can only be
performed by exchanging two successive links. Let
$X=\{X^1,\ldots,X^n\}$, denote the particle types viewed as
letters of an alphabet, and let $\{\lambda^{kl}\}$ be the transition
rates. The set of reactions is defined by
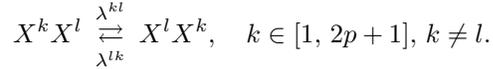
\[
X^k X^l\ \rla_{\lambda^{lk}}^{\lambda^{kl}}\ X^l X^k,\quad
k\in[1,\,2p+1],\, k\ne l.
\]
These rules provide an extension of the $ABC$ model, which we shall
discuss in detail in section \ref{SEC2}.

\medskip
\  \emph{(b)} When $n=2p$ is even, the grammar is altered when two
successive links fold, so that this elementary transition amounts
merely to a rotation of the pleat of angle $\pm\frac{2\pi}{n}$ (instead
of a $\pi$ rotation which would occur when exchanging the two links).
The situation is thus slightly more complicated and  the set of
reactions is now given by
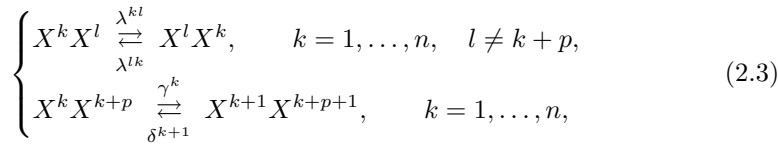
\begin{equation}\label{evnmod}
\begin{cases}
\DD X^k X^l\ \rla_{\lambda^{lk}}^{\lambda^{kl}}\ X^l X^k,\qquad
k=1,\ldots,n, \quad l\ne k+p,\\ 
\DD X^k X^{k+p}\ \rla_{\delta^{k+1}}^{\gamma^k} \ X^{k+1} X^{k+p+1},\qquad
k=1,\ldots,n ,
\end{cases} 
\end{equation}
where $k+p$ is taken modulo $n$. It is worth noting that $\gamma^k$
 (resp. $\delta^k$) concern folds rotating in the counterclockwise
 direction (resp. clockwise) and that the number of letters of each
 type is no longer conserved. In other words, odd models give rise to
 pure diffusions with eventual drifts, when even models are truly
 reaction-diffusion models.

%%%%%%%%%%%%%%%%
\subsection{$3$-dimensional generalizations of 
the coupled exclusion model on a diamond lattice}
\subsubsection{Elementary deformations}

Actually, the diamond lattice formulation of stochastic deformations
in $\mathbb{Z}^3$ provides several straightforward generalization of
the $2$-dimensional $\{\tau_a\tau_b\}$ model. Indeed, between two nodes
there are $8=2^3$ possible links (the jumps of the sample path). Let
$(\tau_a,\tau_b,\tau_c)$ be the vector of binary components
corresponding to a displacement in each direction, where a, b and c
denote here the three particle families (letters).

\medskip
As in the $2$-dimensional model, elementary deformations consist in
exchanging between neighbouring sites the value of one of the binary
components. By construction this model gives a kind of geometric
decoupling between the three types of particles. In fact all possible
existing stochastic coupling result solely from the conditional
transition rates $\lambda_a^{\pm}(\tau_b,\tau_c)$,
$\lambda_b^{\pm}(\tau_c,\tau_a),$ and $\lambda_c^{\pm}(\tau_a,\tau_b)$, according to the various possible models. In order to obtain some non-trivial dynamical effect, we couple the three systems of particles, keeping in mind the possibility of getting
back the $2$-dimensional $\{\tau_a\tau_b\}$ model under certain conditions.
This is the subject of the next two paragraphs.

\subsubsection{A linear coupled exclusion model}
Here we propose a coupling which is linear with respect to the complementary
fields, via the following intensities:
\[
\begin{cases}
\lambda_a^{\pm}(i) = \lambda \pm 2(\tau_i^b - \tau_i^c)\mu_a , \\[0.2cm]
\lambda_b^{\pm}(i) = \lambda \pm 2(\tau_i^c - \tau_i^a)\mu_b ,\\[0.2cm]
\lambda_c^{\pm}(i) = \lambda \pm 2(\tau_i^a - \tau_i^b)\mu_c.
\end{cases}
\]

Suppose for a while $\mu_c=0$ and $\mu_a=\mu_b=\mu$. Then the sequence
$\{tau_c\}$ remains disordered, which means the marginal law of $\tau_i^c$ is
$\frac{1}{2}$. As for the subsystem $\{\tau_a,\tau_b\}$, up
to random contributions $(\tau_i^c-\frac{1}{2})\mu$, the transition rates
$\lambda_a^\pm$ and $\lambda_b^\pm$ correspond to a particular
definition of the $\{\tau_a\tau_b\}$ model. Hence, in the limit, we  obtain a
kind of \emph{disordered} $\{\tau_a\tau_b\}$ model.

\subsubsection{A non-linear coupled exclusion model}
It is also expected to recover the $\tau_a\tau_b$ model when one of the
components is completely frozen in an ordered phase, taking for
example $\tau_i^c=0$ for $i=1,\ldots,N$. This
situation is fulfilled by choosing the following non-linear
couplings
\[
\begin{cases}
\lambda_a^{\pm}(i) = \lambda \pm (\tau_i^b -
\bar\tau_i^b)(\tau_i^c-\bar\tau_i^c)\mu_a,\\[0.2cm] 
\lambda_b^{\pm}(i) = \lambda \pm (\tau_i^c -
\bar\tau_i^c)(\tau_i^a-\bar\tau_i^a)\mu_b ,\\[0.2cm] 
\lambda_c^{\pm}(i) = \lambda \pm (\tau_i^a -
\bar\tau_i^a)(\tau_i^b-\bar\tau_i^b)\mu_c.
\end{cases}
\]

%%%%%%%%%%%%%%%%%
\subsection{Boundary conditions}

Although they live in $\mathbb{Z}^2$ or $\mathbb{Z}^3$, let us
emphasize that all the objects considered throughout this study are
curves, hence one-dimensional dynamical systems.

Hence, for any given sample path of size $N$, there are two links
referred to as site $1$ and $N$, at which \emph{boundary conditions}
have to be specified. We shall consider only \emph{periodic} boundary
conditions: this means essentially the system is \emph{invariant}
under circular permutation of the sites. Consequently, certain
geometric quantities  locally conserved will remain also
globally conserved. For example, the distance between the two
extremities (not necessarily distinct) of a curve remains constant, so
that closed curves will stay closed for ever.

%%%%%%%%%%%%%%%%%%%%%%%%%%%%%%%%%%%%%%%%
\section{Reversibility, Gibbs measures and equilibrium states}\label{SEC2}
In the sequel, our goal will be to find exact scalings permitting to
derive phase transition conditions, for deformation processes of the
class defined in section \ref{SEC1}, as $N\to\infty$. In fact, for the
sake of shortness, we shall restrict ourselves to case studies where
the sample paths have a \emph{Gibbs invariant measure}, but this is by
no means a necessary constraint, as commented in section \ref{CONC}.

\medskip Consider a system having a state space of the form $\mathcal{W}=K^S$,
where $K$and $S$ are finite sets. The following notion, adapted
from \cite{Li}, will be sufficient for our purpose.

\begin{defi}
For any finite set $R\in S$, let $\{V_R\}$ be a collection of real numbers.
A probability measure $\zeta$ on $\mathcal{W}$ is said to be a
\emph{Gibbs state} or a \emph{Gibbs measure} relative to the potential
$\{V_R\}$ if , for all $w=(w_1,\ldots,w_{|S|})\in\mathcal{W}$,
\[
\zeta(w) = \frac{1}{Z} \exp\biggl[\sum_R V_R \prod_{i\in R} w_i\biggr],
 \]
where $Z$ is a normalizing constant.
\end{defi}
 Conditions will now be established, which are either of a geometrical
 nature or bear directly on the transition rates, for the processes of
 interest to be \emph{reversible}: detailed balance equations hold and
 they indeed suffice to equilibrate all possible cycles in the state
 space (see e.g. \cite{Kel}). In that case a potential does exist and
 the invariant measure can be explicitly expressed as a Gibbs measure.

%%%%%%%%%%%%%%%%%%%
\subsection{Odd alphabet}
For $2$-dimensional walks, when the size $n$ of the alphabet is odd,
as already noticed in section \ref{MODELS}, there is a dual point of
view saying that a $n$-species particle system moves in a
one-dimensional lattice: there is exactly one particle per site and
the transition rates $\lambda^{kl}$ correspond to exchanges of a
particle $k$ with a particle $l$ between adjacent sites. In a very
different context, this model was proposed in \cite{EvKaKoMu}, from
which we extract some results pertaining to our topic.

Up to a slight abuse in the notation, we let $X_i^k\in\{0,1\}$
denote the binary random variable representing the occupation of site
$i$ by a letter of type $k$. The state of the system is 
represented by the array $\mathbb{X}\egaldef\{X_i^k, i=1,\ldots,N;
k=1,\ldots,n\}$ of size $N\times\,n$. Then the invariant measure of
the associated Markov process is given by
\begin{equation}\label{inv}
P\bigl(\mathbb{X}\bigr) = 
\frac{1}{Z}\exp\bigl[-\mathcal{H}\bigl(\mathbb{X}\bigr)\bigr],
\end{equation}
where
\begin{equation}\label{ansatz}
\mathcal{H}\bigl(\mathbb{X}\bigr) = 
\sum_{i<j}\sum_{k,l} \alpha^{kl} X_i^k X_j^l,
\end{equation}
$Z$ is the normalizing constant, and
\[
\alpha^{kl}-\alpha^{lk} = \log\frac{\lambda^{kl}}{\lambda^{lk}},
\]
provided that the following condition holds
\begin{equation}\label{oddinv}
\sum_{k\ne l} \alpha^{kl}N_k = 0.
\end{equation}
Indeed, a typical balance equation reads
\begin{equation}\label{balance}
\frac{P[\ldots, X_i^k=1,X_{i+1}^l=1,\ldots]}{P[\ldots,
X_i^l=1,X_{i+1}^k=1,\ldots]} = \frac{\lambda^{lk}}{\lambda^{kl}}=
\exp(\alpha^{lk}-\alpha^{kl}),
\end{equation}
and relation (\ref{oddinv}) proceeds directly from enforcing the above
measure to be invariant by translation.

%%%%%%%%%%%%%%%
\subsection{Even alphabet} 
\begin{figure}[htb]
\begin{center}
\resizebox*{!}{7cm}{\input{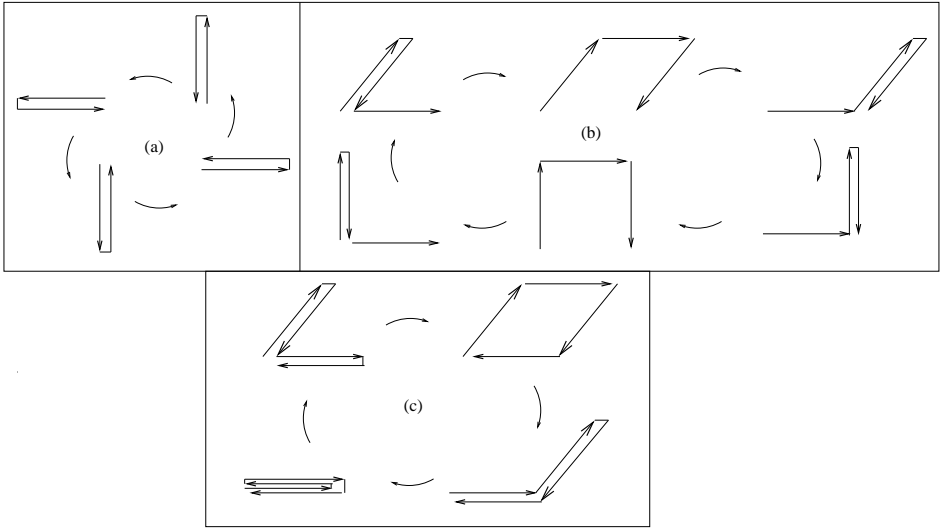}}
\caption{\label{cycles} Elementary cycles: fold (a), 3-link motion (b), 
square loop (c).}
\end{center}
\end{figure}

When the cardinal of the alphabet is even, say $n\egaldef2p$, the
situation is rendered a bit more involved due possible rotations of
consecutive folded links. There is no longer conservation law for the number of letters in each class, and one should instead introduce the quantities
\[
\Delta^k \egaldef N^{k+p} - N^k, \quad k=1,\ldots, p-1, 
\]
which represent the differences between populations of links with
opposite directions. Moreover, as a rule, some non-trivial cycles in
the state-space are not balanced (see figure \ref{cycles}), unless
transition rates satisfy additional conditions. This gives rise to
 the next theorem.
\begin{theo}\label{eventheo}
Assume $n=2p$ and periodic boundary conditions. Then the system is
 \emph{reversible} if and only if the following conditions are
 imposed on the rates and on the particles numbers:
\begin{eqnarray*}
&(i)& \prod_{k=l}^{l+p-1}\frac{\gamma^{k}}{\delta^{k+1}} = 1, \forall
l=1,\ldots,n \\[0.2cm] &(ii)&
\frac{\lambda^{kl}}{\lambda^{lk}}\frac{\lambda^{k+p,\,l}}
{\lambda^{l,\,k+p}} = 1, \quad \forall k,l=1\ldots n, \, k\ne
l,\\[0.2cm] &(iii)& \sum_{l\ne k+p}
\Delta^l\log\frac{\lambda^{kl}}{\lambda^{lk}} = 0, \quad k=1,\ldots,n.
\end{eqnarray*}
\end{theo}

The result relies on  the next lemma.
\begin{lem}
In the case of periodic boundary conditions, if the invariant measure
has a Gibbs form given by (\ref{inv}) and (\ref{ansatz}), then the
following relationships must hold:
\begin{eqnarray*}\label{cond+}
&(iv)& \begin{cases}\DD\alpha^{kl}-\alpha^{lk} =
 \log\frac{\lambda^{kl}}{\lambda^{lk}},\quad k=1,\ldots, n,\ l\ne
 k+p\,; \\[0.2cm] 
\DD\alpha^{k+1,\,k+p+1} - \alpha^{k,\,k+p} =
 \log\frac{\gamma^k}{\delta^{k+1}}, \quad k=1,\ldots, n \,;
\end{cases} \\[0.2cm]
&(v)& \textrm{there exists a constant $\alpha\in\mathbb{R}$ such
that}\\ && \alpha^{kl}+\alpha^{k+p,l}=\alpha^{lk}+\alpha^{l,k+p} =
\alpha, \ \forall k,l = 1,\ldots,n\,; \\[0.2cm] &(vi)&
\sum_{l=1}^{p-1} (\alpha^{kl}-\alpha^{lk})\Delta^l =0, \quad
k=1,\ldots,n.
\end{eqnarray*}
\end{lem}
\begin{proof} We only present the main steps. Condition $(iv)$ 
in the lemma comes from a balance equation of type (\ref{balance}).
The case $l=k+p$ corresponds to adjacent folded links, and, after
setting
\begin{eqnarray*}
U(k,i) & \egaldef &\sum_l\sum_{j>i+1}\bigl(\alpha^{k+1,\,l} +
\alpha^{k+p+1,\,l}\ -\ \alpha^{kl}-\alpha^{k+p,\,l} \bigr)X_j^l , \\
V(k,i) & \egaldef &\sum_l\sum_{j<i}(\alpha^{l,\,k+1} +
\alpha^{l,\,k+p+1}\ -\ \alpha^{lk} - \alpha^{l,\, k+p} )X_j^l,
\end{eqnarray*}
equation  (\ref{balance}) has to be replaced by
\begin{equation*}
\begin{split}
\frac{P[\ldots, X_i^k=1,X_{i+1}^{k+p}=1,\ldots]}{P[\ldots,
X_i^{k+1}=1,X_{i+1}^{k+p+1}=1,\ldots]} & =
\frac{\gamma^k}{\delta^{k+1}}\\[0.2cm] &=
\exp\bigl[\alpha^{k+p+1,\,k+1}-\alpha^{k,\,k+p} - U(k,i) - V(k,i)
\bigr],
\end{split}
\end{equation*}
which leads  to impose condition $(v)$.

To take into account the invariance by translation, let $\sigma$
denote a circular permutation $\sigma$ among the indices, such that
$\sigma(i)=1+i\mod(N)$. Consider
\[
\mathcal{H}_\sigma(\mathbb{X}) = \sum_{1\ge k,l\ge N}\sum_{i<j}
\alpha^{kl} X_{\sigma(i)}^k X_{\sigma(j)}^l,
\]
the resulting energy obtained after applying permutation $\sigma$.
Then
\[
\mathcal{H}_\sigma(\mathbb{X}) =\mathcal{H}(\mathbb{X}) +\sum_{k=1}^N
X_1^k\sum_{l=1}^N (\alpha^{kl}-\alpha^{lk})N^l,
\]
where $N^l$ is the number of links $l$. Since $N^{l+p}-N^{l}$,
$l=1,\ldots,p$ is conserved (but $N^l$ is not), the rule
$(v)$ leads to the  translational invariance in the
form of $(vi)$. 

As to reversibility, one can check that $(i)$ and $(ii)$ are necessary
to equilibrate the cycles depicted in figure \ref{cycles}. Moreover,
the cycle condition imposed by a circular permutation is exactly given
by $(iii)$. Indeed, this cycle is performed by transporting one
particle through the system from site $1$ to site $N$: during this
operation, a tagged particle $X^k$ will encounter the $N^l$ particles
corresponding to all other species $l\ne k$ and the resulting
transition weight is then given by $\DD \prod_{l\ne
k,k+p}\biggl[\frac{\lambda^{kl}}{\lambda^{lk}}\biggr]^{N^l}$, which in
turn, by using $(ii)$, amounts to condition $(iii)$ after taking the
logarithm. These three conditions are in fact sufficient to determine the
parameters $\{\alpha^{kl}\}$ in order to solve $(iv),(v),(vi)$, thus
ensuring reversibility.

\end{proof}

%%%%%%%%%%%%%%%%%%%%%%%%%%%%%%%%%%%%%%%%%%%
\section{Steady state of the ABC system in the thermodynamic limit}\label{SEC3}
As an illustration, we will present a detailed analysis of the
thermodynamic equilibrium situation in the case of the ABC model. When there are 
 three particle species, the form $\mathcal{H}$ of (\ref{ansatz}) comes to
\begin{equation}\label{Hamilton}
\mathcal{H}(\{\mathbb{X}\}) = \sum_{i<j} \alpha^{ab} A_i B_j +
\alpha^{bc} B_i C_j + \alpha^{ca} C_i A_j ,
\end{equation}
where the constants $\alpha^{ab},\alpha^{bc},\alpha^{ca}$ take the
values
\[
\alpha^{ab} = \log\frac{p^+}{p^-},\quad
\alpha^{bc} = \log\frac{q^+}{q^-},\quad
\alpha^{ca} = \log\frac{r^+}{r^-},
\]
and the constraints (\ref{oddinv}) now become
\begin{equation}\label{conditions}
\frac{N_A}{N_B}  = \frac{\alpha^{bc}}{\alpha^{ca}},\quad
\frac{N_B}{N_C} = \frac{\alpha^{ca}}{\alpha^{ab}},\quad
\frac{N_C}{N_A}  = \frac{\alpha^{ab}}{\alpha^{bc}}. 
\end{equation}

%%%%%%%%%%%%%%%%%%%%%%%%%%%%%%%%%%%%%%
\subsection{Scaling and Lotka-Volterra equations}
In the example of the ABC model \cite{ClDeEv}, we have at hand an
explicit analytic expression for the invariant measure. In fact, our
claim is that equations at steady state in the thermodynamic limit can be derived by
using a method proposed in \cite{FaFu}  for the square lattice model,
which a priori \emph{does not require any explicit knowledge} of the
invariant measure, which is most of the time untractable. 

\medskip
We shall  only sketch the main lines of argument in the case of the ABC model.  The number of sites $N$ is fixed and we are interested in the steady state behaviour as $t\to\infty$. 

\medskip 
For the sake of shortness, let $[A,B,C]\egaldef\{(A_i,B_i,C_i),i=1,\ldots,N\}$ denote the $3N$-dimensional bolean vector representing the state of the Markov process.  Combining conditionings and couplings, the approach relies on the construction of a stochastic  \emph{iterative rocking-scheme}, to generate the global invariant measure of the system. 

\medskip 
The algorithm does construct a convergent sequence of random kernels $Q^{(n)},n\ge0$.
\begin{itemize} 
\item[\textbf{S1}]  Set  $n=0$ and take the system in some fixed state  $[A,B,C]^{(0)}=[A,B,C]$.
\item[\textbf{S2}]  At step $n$, assume we are given  a  $3$-tuple $[A,B,C]^{(n)}$, corresponding to an admissible configuration (with respect to the ABC model). To construct $[A,B,C]^{(n+1)}$, one allows each particle family  to evolve as an \emph{independent} one-dimensional exclusion processe until reaching equilibrium, with conditional transition rates \emph{compatible} with the rates of the original process introduced in section \ref{MODELS}. For instance, with regard to  type  $A$ particles \emph{(omitting sometimes the indexing by $n$ for the  sake of brevity)}, we have
\[
\begin{cases}
\lambda_a^+(i) = p^+ B_i + r^- C_i + \Gamma\,\overline
{B}_i\overline {C}_i, \\[0.2cm] \lambda_a^-(i+1) = p^- B_i + r^+ C_i +
\Gamma\,\overline {B}_i\overline{C}_i ,
\end{cases}
\]
these  expressions being similar to the form of (\ref{taux}) in the
$\{\tau_a\tau_b\}$ model. $\Gamma$ is a non-zero quantity taking into
account the exclusion at site $i$ in the following sense: when
$\overline{B}_i\overline {C}_i=1$, then necessarily $A_i=1$. 
In the free process, letting $q_i^a$ denote the random variable representing
the conditional probability of finding a particle $A_i$ at site $i$, we have
\[
\frac{u_{i+1}^{(n)}} {u_i^{(n)}} = \frac{\lambda_a^+(i)}{\lambda_a^-(i+1)},  \quad \mathrm{with} \  u_i^{(n)} \egaldef \frac{q^a_i}{1- q^a_i}.
\] 
Using the  boolean nature of $B_i$ and $C_i$, we can write
\begin{equation}\label{equcond}
\log\frac{u_{i+1}^{(n)}}{u_i^{(n)}} = \alpha^{ca}C_i^{(n)} \, -\, \alpha^{ab}B_i^{(n)} ,
\end{equation}
with analogous quantities $v_i^{(n)}$ and $w_i^{(n)}$ for $B$ and $C$ species respectively.
\item[\textbf{S3}] Clearly, the equilibrium states thus obtained are not all admissible, since at a site $i$,  several particles of different types may coexist. Therefore, we only retain configurations satisfying the constraint 
$A^{(n+1)}_i + B^{(n+1)}_i+ C^{(n+1)}_i= 1$, for all $i$, the distribution of which, after some algebra, takes the form
\[
Q\bigl([(A, B,C)]^{(n+1)} \mid [(A, B,C)]^{(n)}\bigr) = \frac{1}{Z^{(n)}}\prod_{i=1}^N \bigl[u_i^{(n)} A_i ^{(n+1)} + v_i^{(n)} B_i ^{(n+1)} + w_i^{(n)} C_i^{(n+1)} \bigr] ,
\]
where  $Z^{(n)}$ is a normalizing constant.  Then  \textbf{do} $n\leftarrow n+1$ and  \textbf{go to} \textbf{S2}.
\end{itemize}
Let $r^{(n)}_i \egaldef u_i^{(n)}+v_i^{(n)}+w_i^{(n)}$. Then, by means of transfer and coupling theorems, one can prove the existence of the random vectors
 \[
 (r^a_i , r^b_i, r^c_i) = \lim_{n\to\infty} \biggl(\frac{u_i^{(n)}}{r^{(n)}_i},
\frac{v_i^{(n)}}{r^{(n)}_i}, \frac{w_i^{(n)}}{r^{(n)}_i}\biggr) , \quad i\ge 1.
\]
%%%%%%%%%%%%%%%%%%%%
\medskip
\textbf{Fundamental scaling} \  We introduce the so-called
\emph{fundamental scaling} defined by
\[
\alpha^{bc} = \frac{\alpha}{N}+o\Bigl(\frac{1}{N}\Bigr),\quad 
\alpha^{ca} = \frac{\beta}{N}+o\Bigl(\frac{1}{N}\Bigr),\quad 
\alpha^{ab} = \frac{\gamma}{N}+o\Bigl(\frac{1}{N}\Bigr),
\]
 where $\alpha,\beta,\gamma$ are three positive real constant.
 
 \medskip
\textbf{Limiting equations}\  Set from now on $x \egaldef i/N$, for
$1\le i\le N$. Then one can show the weak limits
\[
\rho^a(x) = \lim_{N\to\infty} r_{xN}^a , \qquad \rho^b(x) =
\lim_{N\to\infty} r_{xN}^b , \qquad\rho^c(x) = \lim_{N\to\infty}
r_{xN}^c 
\]
exist and satisfy the system of deterministic differential equations
\begin{equation}\label{Lotka}
\begin{cases}
\DD\frac{\partial\rho_a}{\partial x} = \rho_a(\beta\rho_c -
\gamma\rho_b), \\[0.2cm] \DD\frac{\partial\rho_b}{\partial x} =
\rho_b(\gamma\rho_a - \alpha\rho_c), \\[0.2cm]
\DD\frac{\partial\rho_c}{\partial x} = \rho_c(\alpha\rho_b -
\beta\rho_a),
\end{cases}
\end{equation}
with the  crucial constraints due to the periodic boundary conditions
\begin{equation}\label{boundary}
\rho_u(x+1)=\rho_u(x), \quad \forall u=a,b,c.
\end{equation}
\begin{proof}
Starting from (\ref{equcond}), one applies the law of large numbers
and ergodic theorems to justify the approximation of finite sums by
Riemann integrals, as $N\to\infty$.
\end{proof}

It is amusing to see that (\ref{Lotka}) belongs to the class of
 generalized Lotka-Volterra systems. The original Lotka-Volterra model
 was the simplest model of predator-prey interactions, proposed
 independently by Lotka (1925) and Volterra (1926), see for instance
 \cite{MUR}. Nonetheless, in our case the world is less crual and all
 particle types are  treated on an equal footing...!

\medskip Now the form of (\ref{Lotka}) lends itself to a reasonably explicit  solution in terms of special functions. The first step is to remark the existence of two level
 surfaces
\begin{eqnarray}
\rho_a+\rho_b+\rho_c &= &1 \label{surface1},\\
\rho_a^\alpha\rho_b^\beta\rho_c^\gamma &=& \kappa, \label{surface2}
\end{eqnarray}
where (\ref{surface1}) follows at once from (\ref{conditions}) and
$\kappa$ is a constant of motion to be determined. Using
(\ref{surface1}) to eliminate $\rho_c$, we rewrite (\ref{surface2}) as
\begin{equation}\label{surface3}
\rho_a^{\alpha}\rho_b^{\beta}(1-\rho_a-\rho_b)^{\gamma} = \kappa. 
\end{equation}
The change of functions $u=\rho_a+\rho_b$ and
$v=\beta\rho_a-\alpha\rho_b$, yields the first order nonlinear
differential equation
\begin{equation}\label{diff}
\frac{du}{dx}\ = (1-u)v(u) ,
\end{equation}
where $v(u)$ satisfies the equation 
\begin{equation}\label{polynomial}
(\alpha u + v)^{\alpha} (\beta u - v)^{\beta} (1 - u)^{\gamma} =
\kappa(\alpha+\beta)^{\alpha+\beta} .
\end{equation}

Formally, $u(x)$ can be expressed as
\begin{equation}\label{elliptic}
 x = \int_{u(0)}^{u(x)}\frac{du}{(1-u) v(u)}.
\end{equation} 

It appears that the ratios in (\ref{conditions}) are rational for all
finite $N$, but, since we have let $N,N_A, N_B,\to\infty$, they might
become arbitary real numbers in the interval $[0,1]$. When they are
rational, (\ref{diff}) is a polynomial equation, and then $u(x)$ is in
some sense just a bit more general that an hyperelliptic function,
since, after some algebra, we are left with integrals of the form
\[
\int \sqrt{\bigl[s^2 - a^2(1-s)^{-\frac{p}{q}}\bigr]} ds ,
\]
where $p$ and $q$ stand for positive integers. The particular case
$\alpha=\beta=\gamma$ is rather simple, since then
\[
\left(\frac{du}{dx}\right)^2 = (1-u)[u^2(1-u)-\kappa],
\]
showing $u$ to be a standard Jacobi elliptic function.

\medskip For any given $\kappa$, the system (\ref{Lotka}) admits of
a unique solution. In particular, there is \emph{always} a degenerate
solution crumbled to the fixed point
\begin{equation}\label{fixed}
 \widetilde{\rho}_a = \frac{\alpha}{\alpha+\beta+\gamma}, \quad
 \widetilde{\rho}_b = \frac{\beta}{\alpha+\beta+\gamma}, \quad
 \widetilde{\rho}_c = \frac{\gamma}{\alpha+\beta+\gamma},
\end{equation}
and corresponding to the constant
\begin{equation}\label{constant}
\widetilde{\kappa} \egaldef
\frac{\alpha^{\alpha}\beta^{\beta}\gamma^{\gamma}}
{(\alpha+\beta+\gamma)^{\alpha+\beta+\gamma}}.
\end{equation}
The purpose of the next paragraph is to discriminate between solutions
of (\ref{Lotka}) and (\ref{boundary}), in order to relate them to
admissible limit-points of $\Phi_N$, as $N\to\infty$.

%%%%%%%%%%%%%%%%%%%%%
\subsection{Stability, fundamental period and phase transition}
To catch a rough qualitative insight into the solution of
(\ref{Lotka}), a standard approach relies on a linearization of the
right-hand side around the fixed point (\ref{fixed}). This yields a
linear differential system, whose matrix
\[
\frac{1}{\alpha+\beta+\gamma}
\left[\begin{matrix}
& 0 & -\alpha\gamma & \alpha\beta \\
& \beta\gamma & 0 & -\alpha\beta \\
& -\beta\gamma & \alpha\gamma & 0 \\
\end{matrix}\right],
\]
has three eigenvalues $0$ and $\lambda_\pm= \pm
i\sqrt{\alpha\beta\gamma(\alpha+\beta+\gamma)}$, and the trajectories
are located on an ellipsoid. But, since the above eigenvalues are
purely imaginary, it is well known that no conclusion can be drawn as
for the original system, which might be of a quite different nature. However, it is pleasant to see that the modulus of the nonzero
eigenvalues  plays in fact a crucial role as shown in the next theorem.

\begin{theo}\label{orbit}
Let $s\egaldef\alpha+\beta+\gamma$ and $\DD\eta\egaldef\frac{s}{3}$.
The limit $\DD\Phi\egaldef\lim_{N\to\infty}\Phi_N$ of the ABC model is
deterministic and there is a second order phase transition phenomenon.
There exits a critical value
\[
\eta_c \egaldef \frac{2\pi}
{3\sqrt{\widetilde\rho_a\widetilde\rho_b\widetilde\rho_c}},
\]
such that if $\eta>\eta_c$ then there are closed  non-degenerate 
trajectories of (\ref{Lotka}) satisfying
(\ref{boundary}), with period $T(\kappa_p)=\frac{1}{p}$,
 $p\in\{1,\ldots,[\frac{\eta}{\eta_c}]\}$.
The only admissible stable $\Phi$ corresponds  
\begin{itemize}
\item[--] either to the trajectory associated with $\kappa_1$ if $\eta>\eta_c$;
\item[--] or to the degenerate one consisting of the single point
(\ref{fixed}) if $\eta\le\eta_c$.
\end{itemize}
\end{theo}

The proof involves a forest of technicalities and we only sketch the
main lines of argument. The first step is to switch to polar
coordinates
\[
\begin{cases}
\DD u_a \egaldef \widetilde\rho_a -\frac{\alpha}{s} = r\cos\theta ,
\\[0.2cm] \DD u_b \egaldef \widetilde\rho_b -\frac{\beta}{s} = r\sin\theta.
\end{cases}
\]
Rewrite (\ref{surface2}) as 
\begin{equation}\label{surface4}
H(r,\theta)=\log\kappa,
\end{equation}
with
\[
H(r,\theta)\egaldef\alpha\log\bigl[r\cos\theta + \frac{\alpha}{s}\bigr] +
\beta\log\bigl[r\sin\theta+\frac{\beta}{s}\bigr] +
\gamma\log\bigl[\frac{\gamma}{s}-r(\cos\theta+\sin\theta)\bigr],
\]
and let $r(\theta,\kappa)$ be the single root in $r$ of
(\ref{surface4}). Then $\theta$ satisfies the differential equation
\begin{equation}\label{thetadiff}
\frac{d\theta}{dx} = G(\theta,\kappa),
\end{equation}
where
\begin{eqnarray*}
G(\theta,\kappa) &\egaldef&
 \frac{1}{s}\bigl[\beta(\alpha+\gamma)\cos\theta +
 \alpha\beta\sin^2\theta +
 \alpha(\beta+\gamma)\sin^2\theta\bigr]\\[0.2cm] &+&
 r(\theta,\kappa)\cos\theta\sin\theta\bigl[(\beta+2\alpha+2\gamma)
 \cos\theta + (\alpha+2\beta+2\gamma)\sin\theta\bigr].
\end{eqnarray*}

Letting $T(\kappa)$ be the period of the orbit,  we have
\[
T(\kappa) = \int_0^{2\pi} \frac{d\theta}{G(\theta,\kappa)}.
\]
The second important step relies on the monotonic behaviour of
 $T(\kappa)$ with respect to the parameter $\kappa$,  yielding the
 inequality
\[T(\kappa)\ge T(\widetilde\kappa).
\] 
Observing that 
\[
r(\theta,\widetilde\kappa) = 0,\quad\forall\,\theta\in[0,2\pi],  
\]
we can write, by (\ref{thetadiff}), $T(\widetilde\kappa)$ as a contour
integral on the unit circle, namely
\[
T(\widetilde\kappa) = -4is\oint_\Gamma
\frac{1}{z^2[\gamma(\beta-\alpha)-2i\alpha\beta]+2z[2\alpha\beta +
\gamma(\alpha+\beta)]+\gamma(\beta-\alpha)+2i\alpha\beta},
\]
or, after a simple calculus, 
\[
T(\widetilde\kappa) = 2\pi\sqrt{\frac{s}{\alpha\beta\gamma}},
\]
which leads precisely to  the critical value $\eta_c$ announced in the theorem.

%%%%%%%%%%%%%%%%%%%%%%%%%%%%%%%%%%%%%%%%%%%

\section{Perspectives}\label{CONC}
This paper is the continuation of \cite{FaFu}, but is certainly an
intermediate step. For the sake of shortness, we did restrict
ourselves to the thermodynamics of the ABC model. Actually, our goal
is to analyze the dynamics of random curves evolving in $\mathbb{Z}^m$
(no spatial constraints) or in $\mathbb{Z}_+^m$, when they warp under
the action of some stochastic deformation grammar.

In \cite{FaFuII}, this project will be carried out in the framework of
large scale dynamics for exclusion processes, and it will mainly
address the points listed hereafter.

\textbf{More on thermodynamic equilibrium} \quad The trick to derive
limiting differential systems amounts essentially to writing
conditional flow equations on suitable sample paths, even in the
presence of particle currents. These equations involve functionals of
Markov and they enjoy special features encountered in many systems. It
might also be interesting to note that most of the Lotka-Volterra
equations can be explained in the light of the famous urns of
Ehrenfest.

\textbf{Phase transition}\quad There exists a global interpretation by
means of a free energy functional with two components: the entropy of
the system, and the algebraic area enclosed by the curve. It turns out
that the contention between these two quantities yield, after taking
limits $\lim_t\to\infty\lim_N\to\infty$ (in that order), either
stretched deterministic curves or Brownian objects when the scaling is
of central limit-type.

\textbf{Transient regime} \quad Our claim is that
time-dependent behaviour can be treated along the same ideas, up to
technical subtleties, by means of a numerical scheme based on the
conservation of particle currents. This should yield a system of Burgers
equations, extending those obtained in \cite{FaFu} for the symmetric
$\{\tau_a\tau_b\}$ model, which had the form
\begin{equation*} 
\begin{cases}
\DD\frac{\partial\rho^a(x,t)}{\partial
 t} \ = \ D\frac{\partial^2\rho^a(x,t)}{\partial x^2}
 -2D\eta\frac{\partial}{\partial
 x}\bigl[\rho^a(1-\rho^a)(1-2\rho^b)\bigr](x,t), \\[0.2cm]
\DD \frac{\partial\rho^b(x,t)}{\partial
 t} \ = \ D\frac{\partial^2\rho^b(x,t)}{\partial x^2}
 +2D\eta\frac{\partial}{\partial
 x}\bigl[\rho^b(1-\rho^b)(1-2\rho^a)\bigr](x,t). 
\end{cases}
\end{equation*}

\bibliography{resume}

\begin{thebibliography}{99}

\bibitem{Bu}
{\sc J.~Burgers}, {\em A mathematical model illustrating the theory of
  turbulences}, Adv. Appl. Mech., 1 (1948), pp.~171--199.

\bibitem{ClDeEv}
{\sc M.~Clincy, B.~Derrida, and M.~Evans}, {\em Phase transition in the {ABC}
  model}, Phys. Rev. E, 67 (2003), pp.~6115--6133.

\bibitem{DeEvHaPa}
{\sc B.~Derrida, M.~Evans, V.~Hakim, and V.~Pasquier}, {\em Exact solution for
  1d asymmetric exclusion model using a matrix formulation}, J. Phys. A: Math.
  Gen., 26 (1993), pp.~1493--1517.

\bibitem{EvFoGoMu}
{\sc M.~Evans, D.~P. Foster, C.~Godr\`eche, and D.~Mukamel}, {\em Spontaneous
  symmetry breaking in a one dimensional driven diffusive system}, Phys. Rev.
  Lett., 74 (1995), pp.~208--211.

\bibitem{EvKaKoMu}
{\sc M.~Evans, Y.~Kafri, M.~Koduvely, and D.~Mukamel}, {\em Phase {S}eparation
  and {C}oarsening in one-{D}imensional {D}riven {D}iffusive {S}ystems}, Phys.
  Rev. E., 58 (1998), p.~2764.

\bibitem{FaFuII}
{\sc G.~Fayolle and C.~Furtlehner}, {\em Stochastic deformations of random
  walks and exclusion models. {P}art {II}: Gibbs states and dynamics in
  $\mathbb{Z}^2$ and $\mathbb{Z}^3$}.
\newblock In preparation.

\bibitem{FaFu}
{\sc G.~Fayolle and C.~Furtlehner}, {\em Dynamical {W}indings of {R}andom
  {W}alks and {E}xclusion {M}odels. {P}art {I}: Thermodynamic limit 
in $\mathbb{Z}^2$}, Journal of Statistical Physics, 114 (2004), pp.~229--260.

\bibitem{Ka}
{\sc O.~Kallenberg}, {\em Foundations of Modern Probability}, Springer, second
  edition~ed., 2001.

\bibitem{KPZ}
{\sc M.~Kardar, G.~Parisi, and Y.~Zhang}, {\em Dynamic scaling of growing
  interfaces}, Phys. Rev. Lett., 56 (1986), pp.~889--892.

\bibitem{Kel}
{\sc F.~P.~Kelly}, {\em Reversibility and stochastic networks}, John Wiley \&
  Sons Ltd., 1979.
\newblock Wiley Series in Probability and Mathematical Statistics.

\bibitem{LaBaRa}
{\sc R.~Lahiri, M.~Barma, and S.~Ramaswamy}, {\em Strong phase separation in a
  model of sedimenting lattices}, Phys. Rev. E, 61 (2000), pp.~1648--1658.

\bibitem{Li} {\sc T.~M.~Liggett}, {\em Interacting Particle Systems},
vol.~276 of Grundlehren der mathematischen {W}issenschaften,
Springer-Verlag, 1985.

\bibitem{MUR}
{\sc J.~Murray}, {\em Mathematical Biology}, vol.~19 of Biomathematics,
  Springer-Verlag, second~ed., 1993.

\bibitem{Sp}
{\sc H.~Spohn}, {\em Large Scale Dynamics of Interacting Particles}, Springer,
  1991.

\end{thebibliography}

%%%%%%%%%%%%%%%%%%%%%%%%%%
%% Name and Affiliation %%
%%%%%%%%%%%%%%%%%%%%%%%%%%

\add{Guy Fayolle}
{INRIA Rocquencourt -- Domaine de Voluceau BP 105\\
78153 Le Chesnay, France.\\
Guy.Fayolle@inria.fr}
\medskip
\add{Cyril Furtlehner}
{INRIA Rocquencourt -- Domaine de Voluceau BP 105\\
78153 Le Chesnay, France.\\
Cyril.Furtlehner@inria.fr}

\end{document}